\documentclass[sigconf]{acmart}

\def\BibTeX{{\rm B\kern-.05em{\sc i\kern-.025em b}\kern-.08emT\kern-.1667em\lower.7ex\hbox{E}\kern-.125emX}}
\usepackage{algorithmic}
\usepackage{algorithm}   
 
\begin{document}

%
\title{A Deep Learning Approach for Tweet Classification and Rescue Scheduling for Effective Disaster Management (Industrial)\\}

%
\author{Md. Yasin Kabir}
\affiliation{%
 \institution{Department of Computer Science\\ Missouri University of Science and Technology, USA}
}
 \email{mkdv6@mst.edu}

\author{Sanjay Madria}
\affiliation{%
 \institution{Department of Computer Science\\ Missouri University of Science and Technology, USA}
}
  \email{madrias@mst.edu}

%

%
\begin{abstract}
Every activity in disaster management such as managing evacuation plan, and running rescue missions demands accurate and up-to-date information to allow a quick, easy, and cost-effective response to reduce the possible loss of lives and properties. It is a challenging and complex task to acquire information from different regions of a disaster-affected area in a timely fashion. The extensive spread and reach of social media and networks allow people to share information in real-time. However, the processing of social media data and gathering of valuable information require a series of operations such as (1) processing each specific tweet for a text classification, (2) possible location determination of people needing help based on tweets, and (3) priority calculations of rescue tasks based on the classification of tweets. These are three primary challenges in developing an effective rescue scheduling operation using social media data. In this paper, first, we propose a deep learning model combining attention based Bi-directional Long Short-Term Memory (BLSTM) and Convolutional Neural Network (CNN) to classify the tweets under different categories. We use pre-trained crisis word vectors and global vectors for word representation (GLoVe) for capturing semantic meaning from tweets. Next, we perform feature engineering to create an auxiliary feature map which dramatically increases the model accuracy. In our experiments using real data sets from Hurricanes Harvey and Irma, it is observed that our proposed approach performs better compared to other classification methods based on Precision, Recall, F1-score, and Accuracy, and is highly effective to determine the correct priority of a tweet. Furthermore, to evaluate the effectiveness and robustness of the proposed classification model a merged dataset comprises of 4 different datasets from CrisisNLP and another 15 different disasters data from CrisisLex are used. Finally, we develop an adaptive multi-task hybrid scheduling algorithm considering resource constraints to perform an effective rescue scheduling operation considering different rescue priorities.
\end{abstract}



\maketitle

\section{Introduction}
The exponential growth in social media such as Twitter and Facebook experiencing their mass adaptation in several applications. 
The roles of social media extended but not limited to health and disease analysis and propagation detection \cite{park2017social}, 
Quantifying controversial information \cite{garimella2018quantifying}, 
and disaster crisis management \cite{alexander2014social, yang2017harvey}. Natural disasters frequently disrupt regular communication due to the damaged infrastructures \cite{shklovski2010technology} which lead to an outflow of information. A report on Hurricane Sandy \cite{baer2012sandy} shows that people were using social media more frequently to communicate. People were seeking help quickly and promptly as they strive to contact friends and family in and out of the disaster area, looking for information regarding transport, shelter, and food. Hence, the huge flow of information over social media can be beneficial in managing a natural disaster more effectively. During Hurricane Sandy, Twitter proved its usefulness, and at the time of Hurricane Harvey and Irma, again Twitter played a crucial role in the rescue, donation, and recovery. However, while the use of social network seems appealing, still most of the applications 
are still lacking features and fall short in their usability \cite{lindsay2011social}. 
\begin{center} 
    \centerline{\includegraphics[width=3.2in]{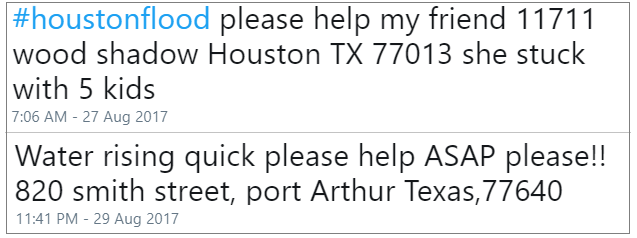}}
    \label{rescue_tweets} 
    \vspace{-1em}
     \captionof{figure}{Examples of rescue requesting tweets}
    \vspace{-0.5em}
\end{center}

\begin{center} 
    \centerline{\includegraphics[width=3.2in]{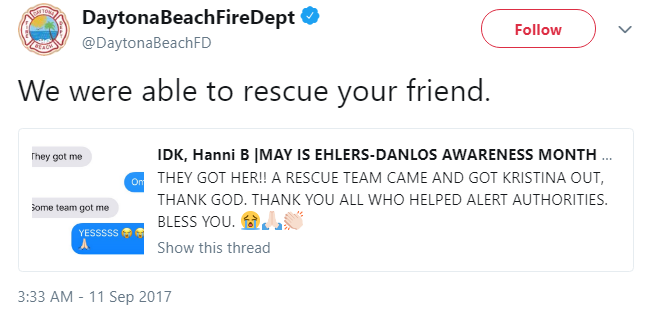}}
    \vspace{-1em}
    \captionof{figure}{A successful rescue during Hurricane Irma}
    \label{rescue_success}
\end{center}
\vspace{-1em}
Figure ~\ref{rescue_tweets} represents two tweets seeking rescue during Hurricane Harvey. People also tweeted similarly at the time of Hurricane Irma. 
Figure ~\ref{rescue_success} shows a tweet after the successful rescue mission during Hurricane Irma. 

Institutional and Volunteer rescue efforts save a lot of lives during a crisis. However, those rescue missions are not well-organized and structured due to uncertainty. Individual volunteers have time constraints and lack of resources. Moreover, some rescue missions might need extra precaution, advanced equipment, and medical facilities. Besides that, due to the variety of help requesting tweets, some of those tweets might be out of sight. Hence, an automated system is essential to understand the context of the tweets, classify the specific tweets for rescue, prioritize those tweets based on context, and then schedule rescue missions and allocate necessary resources accordingly. Our primary contributions in this paper are:

\begin{itemize}
\item Developing a multi-headed binary classifier to classify the tweets into six different classes using deep learning where a single tweet can belong to multiple classes. We use a unique machine learning pipeline with a set of punctuation-based auxiliary features which are specifically correlated with the disaster-related tweets.
\item Evaluating and comparing the proposed model with different machine learning models and diverse datasets.  
\item We formally introduce a method for priority determination of each rescue request which plays a crucial role in maintaining fairness in the rescue scheduling. 
\item We propose a resource constraint and burst time adaptive rescue scheduling algorithm with multi-tasking and priority balancing to perform improved rescue operations. 
\end{itemize}


 
\section{Related Work}
\subsection{Social Media Analysis for Disaster Management}
Most of the prior research work research works using social media and networks for disaster management are focused on assessing the disaster situation, and a little, if any, is focused on their use in rescue mission and planning. Authors in \cite{yang2017harvey} proposed a rescue scheduling algorithm on Hurricane Harvey which connects the victims with the scattered volunteers. A heuristic multi-agent reinforcement learning scheduling algorithm, named as ResQ \cite{nguyen2018coordinating}, 
utilizes reinforcement learning to coordinates the volunteers and the victims during a disaster. \cite{yin2015using} proposed a system that uses machine learning mechanism to extract the data that is generated by Twitter messages during a crisis. 
Authors in \cite{gao2011harnessing} presented a research showing how social media communication was used during the catastrophic Haiti earthquake. 
They adapted the method of crowd-sourcing for designing coordination protocols and mechanisms in order to create coordination between the organizations and their relief activities. \cite{starbird2010pass} analyzed the extensive use of Twitter data in case of mass convergence or disaster situation such as the Southern California Wildfire. 
After several devastating incidents, a few disaster management applications such as Ushahidi \cite{okolloh2009ushahidi} have been developed. 


\subsection{Tweets Classification}
The basic approach for tweet classification is to extract features from the text. 
Naive Bayes classifier\cite{rish2001empirical} is one of the popular classifiers which models the document distribution using probability. Another widely used entity in classification is Support Vector Machines (SVM), which draws a linear separator plane among the classes \cite{allahyari2017brief}. To perform the classification, K-Nearest Neighbor Classifier \cite{han2001text} offers proximity-based classifier, and uses distance measurement among the words. 

The idea of the deep neural network for natural language processing first used in \cite{lapata2017proceedings} uses a multitask learning model using the neural network. \cite{conneau2016very} proposed a deep neural network consisting of 29 layers for natural language processing. 
\cite{imran2016twitter} and \cite{nguyen2016rapid} showed that combining with external pre-learned word vectors such as GloVe \cite{pennington2014glove}, a neural network can be trained better for the disaster datasets. Our proposed deep learning model took inspiration form their work. However, those works did not consider any auxiliary features or attention layer. As a tweet has character length restriction, attention layer with domain-specific engineered auxiliary features can be highly influential. In this work, we create a set of auxiliary features and use an attention based deep neural network to classify the tweets into 6 different classes where each class represents a binary output label, and a single tweet can belong to multiple classes. 


\subsection{Scheduling Algorithms}
The scheduling algorithms intend to optimize the time and the use of resources among different parties employing certain constraints. The primary purpose of a scheduling algorithm is to ensure fairness among the participants while maximizing resource utilization. First-Come-First-Served (FCFS) 
algorithm can not provide fairness when someone cannot wait to use the resource or when someone needs a priority based on a situation. \cite{schwiegeishohn1998improving} improved the FCFS using parallel processing technique. \cite{leung1982complexity} worked with fixed-priority scheduling to consider the complexity of determining whether a set of periodic real-time tasks can be scheduled on $m > 1$. \cite{huang2016self} proposed fixed-priority scheduling using a fixed-relative deadline. After a certain period of time, a task became suspended upon failure and the resource became available. \cite{hu2019scheduling} presents a scheduling algorithm for emergency medical rescue conflict monitoring and dispatch scheduling based on the hybrid estimation and intent inference. 
\cite{wex2014emergency} took a heuristic approach for solving the rescue unit assignment and scheduling problem under the resource constraints. In \cite{yang2017harvey}, the authors discuss the utilization of the public resources for disaster rescue with the priority based scheduling policy. The authors present a discussion about the fairness and importance of priority based on rescue scheduling. 
However, there is no formulation to determine the priority scores of rescue scheduling tasks. In this paper, we formally define a method to determine the priority score of rescue tasks and propose a multi-task hybrid scheduling policy using priority, based on certain criteria to develop an effective and efficient rescue scheduling algorithm. 

\section{Tweets Classification and Information Retrieval}\label{dataColPro}
Twitter data from two different natural disasters (Hurricane Harvey and Hurricane Irma) were collected for this work. We collected these tweets from August 26 to August 31, 2017 and September 10 to September 17, 2017, respectively. We use Twitter Stream API to collect the tweets along with various meta-information such as user information, geo-location, tags, entities, etc. A simple work-flow for Tweets collection and information retrieval is shown in Figure ~\ref{workflow}. The pre-processing step involves discarding non-English tweets, filtering noises and duplicates, removing special characters, stop-words, and jargons. 

\begin{center}
    \includegraphics[width=3.2in]{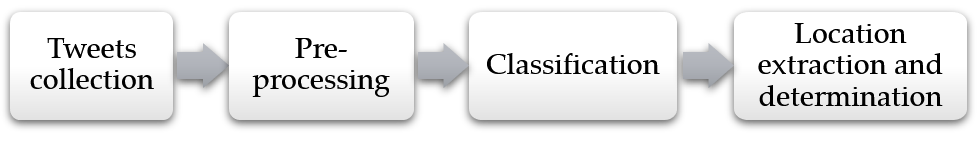}
    \vspace{-1em}
    \captionof{figure}{Simple workflow of data collection and information extraction}
    \label{workflow}
\end{center}

A tweet classifier is developed using the neural network to identify whether a tweet falls into one or more classes from six different classes (Rescue needed, DECW, Water needed, Injured, Sick, Flood). DECW stands for Disabled, Elderly, Children and Women. According to the FEMA \cite{coursesummary, fema2}, WHO \cite{who_2012}, and NCDP \cite{ncdp}, the "vulnerable populations" or "at-risk individuals" includes children, senior citizens, pregnant women, disabled, sick or injured persons. Hence, we labeled and categorized the tweets accordingly. The label "Rescue needed" is the base label which identifies if a tweet is seeking rescue or assistant to evacuate. We use the label "Water Needed" as a request for drinkable water identified as a vital resource during any disaster or emergency by the CDC \cite{cdc}. Finally, we also used the label "Flood" because flood is common during hurricanes and should be considered appropriately for determining the priority in rescuing a victim and preparing the rescue mission with appropriate resources.

Those six classes help in determining the rescue situation, and their priorities along with the resources needed or requested in a tweet. To classify a tweet, the tweet text along with a processed set of auxiliary features fed into the classification algorithm.

\vspace{-1.5em}
\subsection{Deep Neural Network}\label{CNN}
\begin{figure}[ht]
  \includegraphics[width=3.4in]{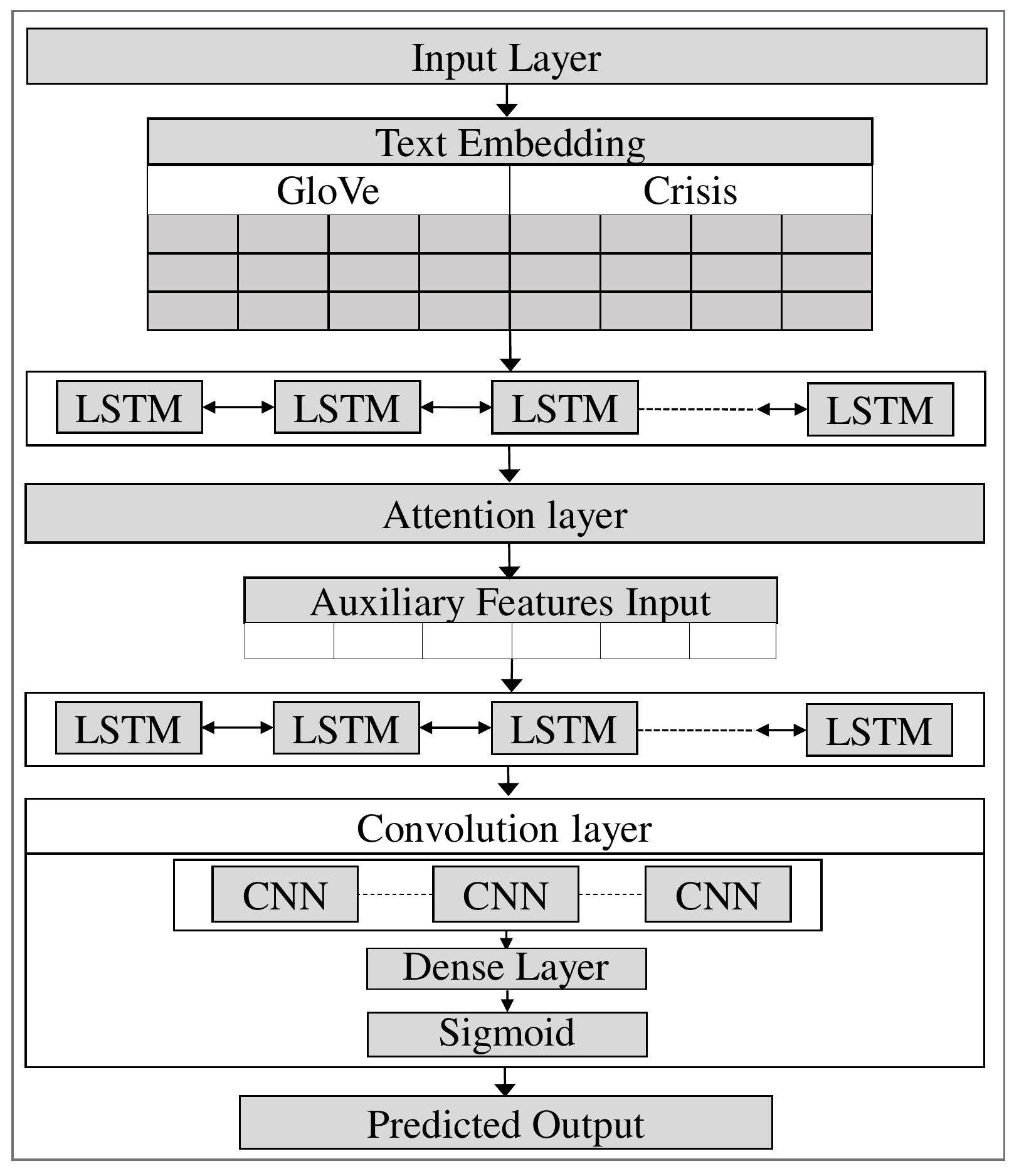}
  \vspace{-1em}
  \caption{System architecture of the proposed model}
  \label{fig:systemArchitecture}
  \vspace{-1em}
\end{figure}

The proposed deep learning model comprises 7 primary components: the Input layer, Embedding layer, BLSTM layer, Attention layer, Auxiliary features input, Convolution layer, and Output. Figure \ref{fig:systemArchitecture} depicts the fundamental system architecture of the model.
\vspace{-1em}

{\textbf{Input layer:}} Pre-processed tweets fed to the input layer which is connected with the embedding layer.

{\textbf{Embedding layer:}} This layer encodes the input into real-valued vectors using lookup tables. Pre-trained word embedding is proven beneficial to understand the semantic meaning of the words and improve the classification models. In this work, we used a pretrained word vectors named Crisis \cite{imran2016twitter} and GloVe \cite{pennington2014glove} which generates a feature word vectors using co-occurrences based statistical model. Embedding applied to the words aids to map all tokenized words in every tweet to their respective word vector tables. To unify the feature vector matrix, appropriate padding is added.

{\textbf{BLSTM layer:}} The Long-Short Term Memory (LSTM) is a specialized version of Recurrent Neural Network (RNN) that is capable of learning long term dependencies. While LSTM can only see and learn from past input data, Bidirectional LSTM (BLSTM) runs input in both forward and backward direction. This bidirectional feature of BLSTM is critical for the various applications involved with understanding complex language \cite{wang2015learning} as it can capture both future and past context of the input sequence. 
\begin{flalign}
  & i_t = \sigma\left(W_{xi}x_t+W_{hi}h_{t-1}+W_{ci}c_{t-1}+b_i \right) \\
  & f_t = \sigma \left( W_{xf}x_t + W_{hf}h_{t-1} +W_{cf}c_{t-1} + b_f \right) \\
  & o_t = \sigma \left( W_{xo}x_t + W_{ho}h_{t-1} +W_{co}c_t + b_o \right) \\
  & c_t = f_tc_{t-1} + i_t\tanh \left( W_{xc}x_t + W_{hc}h_{t-1} + b_c \right) \\
  & h_t = o_t \tanh \left( c_t \right) 
\end{flalign}

The input gate $i_t$, forget gate $f_t$, output gate $o_t$, and cell state activation $c_t$ of the implemented LSTM version in this work can be defined by the equations (1)-(5)  where $\sigma$ represents the logistic sigmoid function, $h$ represents the respective hidden vectors, and $W$ is the weight matrix. A detailed explanation of each equation and more about LSTM are available on \cite{graves2013speech}.

{\textbf{Attention layer:}} Every word in a sentence does not contribute equally to represent the semantic meaning and the primary concept of attention \cite{luong2015effective} originated from this observation. We use a word-level deterministic, differentiable attention mechanism to identify the words with the closer semantic relationship in a tweet. Equation 6 represents the attention score $e{i,e}$ of each word $t$ in a sentence $i$, where $g$ is an activation function. More information on the attention mechanism is available on \cite{kumar2019sarcasm}.
\begin{equation}
e_{i,j} = g \left( Wh_tc\right)
\end{equation}
 
{\textbf{Auxiliary features input:}} A tweet can only contain 280 characters (previously 140) which forces a user to express emotions and opinions in a different way compared to a traditional English sentence. People use extra punctuations and emoticons to intensify the meaning of a tweet. We also observed (e.g. Figure \ref{rescue_tweets}) greater use of numeric characters in a rescue seeking tweet due to the fact that people try to share location in the tweets, which most of the time essentially contains digits. In this work, we perform feature engineering to obtain a set of specific auxiliary features that can assist the classification model to learn better. A list of extracted auxiliary features that shows noticeable influence during the model evaluation is given in Table \ref{aux_features}. The well-known Natural Language Toolkit (NLTK) \cite{loper2002nltk} is used to extract those features. 

\vspace{-0.7em}
\begin{table}[H]
\renewcommand{\arraystretch}{1}
\caption{Auxiliary Features}
\vspace{-1em}
\label{aux_features}
\centering
\begin{tabular}{|p{8cm}|}
\hline
polarity, subjectivity, sentiment, wordsVsLength, exclamationMarks, questionMarks, digitVsLength, digitVsWord, punctuationVsLength, punctuationVsWords, nounsVsWwords, sadVsWords, angryVsWords, capitalsWords, capitalsVsWords, uniqueWords, repeatedWords, numberOfHashtags. \\
\hline
\end{tabular}
\end{table}

{\textbf{Convolution layer}} The convolution layer performs a matrix-vector operation in the sentence-level representation sequence. Let us assume that $H \in \mathbb{R}^{d*w}$ be the weight matrix, and the feature mapping done as $ c \in \mathbb{R}^{l-w+1}$. The i-th element of the feature map can be defined as:
\begin{equation}
\label{cnn_mapping}
c_i = \sigma \left( \sum \left( C \left[ *,i:i+w \right] o  H\right) + b \right) 
\end{equation}
In sentence-level representation, $C[*,i:i+w]$ is the i-th to i+w-th column vector. The word vectors pass through the convolution layers \cite{wang2016combination} where all the input information merged together to produce a features map. The Rectified Linear Unit (ReLU) used as the activation to deal with the non-linearity in the convolution layer and generate a rectified feature map. Finally, the dense layers are activated for generating the outputs.

{\textbf{Output layer}} The activation function $sigmoid$ is used in the dense layer as we want to perform multi headed binary classification. The model produces binary values for all six target output classes. Detailed information on model hyperparameters and evaluation results is given in section \ref{CE}. 


 

\subsection{Location extraction}\label{LE}
Due to the privacy policy of Twitter, most of the tweets do not contain any location information. In those cases, we try to extract location using user profile meta information and the location information provided in the tweeted text. Combining the Stanford Named Entity Recognizer (NER) \cite{finkel2005incorporating} and Google map API, an application is built for extracting location. 

\section{Rescue Scheduling}\label{RSalgos}
\subsection{Problem Specification}\label{PS}
Let us assume that the number of rescue teams be $m$ with $n$ pending rescue tasks. Let the processing time of rescue task $j$ by team $i$ be $t_{ij}$, where $1\leq i \leq m, 1 \leq j \leq n$. Based on a typical disaster situation, we consider that the number of rescue tasks is greater than or equal to the number of rescue teams ($n\geq m$). The problem is to organize and assign the tasks to rescue teams in such a way that the amount of waiting time for each rescue mission is minimized. However, due to the inconsistent nature of the rescue tasks and the location of the incidents, the formulation of this problem faces the following major challenges.

\begin{enumerate}
\item Every rescue team may not capable of processing each task. We need to consider specific requirements of each task and different capabilities of rescue units.   
\item It is difficult to precisely estimate the required time $t_{ij}$ for a task due to the uncertainty of the environment and location of an incident. 
\item Tasks might have different priorities based on the people needed to execute them and their physical condition. The environmental condition of a person such as surrounded by flood water, or fire should also be taken into account while determining priority. 
\end{enumerate}

Along with the above challenges, we also consider the following restrictions and conditions to formulate the problem effectively. 
\begin{itemize}
\item We imposed a time $t_j$ for a task $j$, where $t_j$ denotes the required time for a rescue team $i$ to move from initial rescue center to the place of incident. The time for moving from the location of a task $j$ to another task $j'$ is represented by $t_{jj'}$.
\item Every team requires a preparation time before leaving for a scheduled rescue job from their respective rescue management station. The preparation time is denoted as $t_i$ for every team for a specific task j. Also, after a certain period, every team might require a resting time of $t_{ir}$ before the next task. 
\end{itemize}
Considering the above sequence of times ($t_{ij}$, $t_j$, $t_{jj'}$, $t_i$, and $t_{ir}$), we can estimate a probable time for a rescue mission. Although the time can be changed based on the situation, we consider some constant time variable considering the distance of a task location and the probable situation of the environment around the incident.  

\subsection{Priority determination}\label{PD} 
A significant step for the rescue scheduling algorithm is determining the priority of rescue tasks. We use the output labels of tweet classifier ( Section \ref{CNN}) and assign a weight for each label to determine the priority of that tweet. Assume that the assigned weights for different labels of the tweets is represented by a vector $w_j = [w_{1},w_{2},...w_{n}]$. A feature vector $\alpha_i = [\alpha_{1},\alpha_{2},...\alpha_{m}]$ also used which denotes the weight of other considerable variables such as the number of victims, real-time environmental conditions and future weather forecasts of a specific location. The equation \ref{priority_func} represents the formula to estimate the priority for a rescue task. The base priority value of a tweet is 1 where the maximum priority score can be 10. 
\vspace{-.5em}
\begin{equation}
\label{priority_func}
f_p = \sum\limits_{i=1}^m\alpha_i + \sum\limits_{j=1}^n w_j 
\end{equation}

\subsection{Rescue scheduling algorithms}\label{Algos}
General scheduling algorithms are not applicable in disaster rescue scenario as those algorithms might be unfair due to different situations, physical conditions, and the critical importance of human life. A priority-based scheduling algorithm might provide a better solution where we need to consider and determine the priority continuously. In a disaster scenario, priorities can change with time and environmental conditions. Hence, We develop an effective rescue scheduling algorithm considering priority, environmental severity, and processing time of every single task. We like to define the terms which we use to represent our algorithms. 

\begin{itemize}
\item Tasks: A task is the combination of one or more valid rescue requests by an individual or multiple people. A list of valid requests forms a sequence of tasks which demands to be scheduled appropriately.
\item Processors: The number of rescue units which can complete a given task is the processors. A processor is responsible to execute a given task, release the resources upon completion, and get back to the initial state to execute a new task.
\item Arrival Time: The time of receiving a valid rescue request represents the arrival time for a specific task. In our rescue scheduling system, arrival is the time-stamp of a tweet. 
\item Burst Time: The probable time required to complete a task by a processor can be defined as burst time. The burst time is realistically represents the service time of a processor for a rescue mission. In a disaster scenario, estimating appropriate burst time is very challenging. Similar tasks might take different times to complete under separate circumstances. To address this issue, first, we assume a probable burst time based on the rescue operations in previous disasters. After the completion of a few rescue missions, the burst time of the future mission is determined using the actual completion time of those missions. To predict the future burst time, we use the exponential averaging method. Given n tasks (taskSeq[1...n]) and burst time for tasks $t_i$, the predicted burst time for the next task $taskSeq_n+1$ will be:
\begin{equation}
\label{burstTimePred}
BT_{n+1} = \alpha T_n + (1-\alpha)BT_n
\end{equation}

In the above equation, $\alpha$ is a constant factor ranging ($0<=\alpha<=1$). The value that can predict the best possible burst time will be assigned as $\alpha$. The variable $BT_n$ denotes the predicted or assumed burst time for the task n, and  $T_n$ represents the actual burst time needed for completing task n.
\end{itemize} 

Three different scheduling algorithms are implemented for the experiments. All of those algorithms are implemented using multiple processors as it is expected to have more than one rescue unit in an emergency rescue situation. Although we emphasize on Multi-task Hybrid Scheduling algorithm, however, we study fundamental rescue algorithms to understand the limitations of these established methods. This study also indicates the necessity of a novel adaptive Hybrid Scheduling algorithm for a disaster scenario. 

\subsubsection{First-Come-First-Serve (FCFS)}
In FCFS scheduling system, the task requests are sequentially processed in the order of the arrival time. A sequence of tasks list (taskSeq) with the requests arrival time (arrivalTime) and probable burst time (burstTime) is fed to the algorithm as input. The algorithm returns the scheduled tasks sequence with the possible start time. However, estimate burst time can change and needed to update while the processor is processing a task. While FCFS is a simplest scheduling algorithm, it has two major concerns which need some attention.
\begin{itemize}
\item In a disaster scenario, every rescue request is not similarly critical. FCFS fails to consider the tasks which have an urgency of completion. 

\item FCFS is a non-preemptive scheduling algorithm which is responsible for the short jobs to wait longer based on the sequence order.  
\end{itemize}

\subsubsection{Priority Scheduling} 
In a disaster scenario, conducting rescue missions based on priority can be crucial. There can be lots of rescue requests which can wait longer, and might not be as critical as some other requests. A priority-based scheduling algorithm can be more appropriate considering those facts. The algorithm executes the task based on an ordered queue with high to low priority values. A priority queue based scheduling algorithm is demonstrated in the Algorithm 1.

\setlength{\textfloatsep}{0pt}
\begin{algorithm}
\label{priorityAlgo}
 \caption{Priority scheduling with multi-processors}
 \begin{algorithmic}[1]
 \renewcommand{\algorithmicrequire}{\textbf{Input:}}
 \renewcommand{\algorithmicensure}{\textbf{Output:}}
 \REQUIRE processorNo, taskSeq$[1...n]$, arrivalTime$[at_1...at_n]$, burstTime$[bt_1....bt_n],$ tasksPriority$[1...n]$;
 \ENSURE  scheduleSeq$[task_i...task_n]$, startTime$[st_1...st_n]$, \\ turnAroundTime, averageWaitingTime, averageTurnAroundTime;
 \\ \textbf{Initialization}: All the processors K are released and ready to begin a task.\\ Initialize, scheduleSeq, startTime, and turnAroundTime as list []; currentTime = 0, waitingTime = 0, totalTurnAroundTime = 0;\\
Sort the taskSeq, arrivalTime, burstTime using taskPriority and assign the tasks in priority queue $P_{queue}$;
 \IF {(new task request)}
 \STATE update $P_{queue}$, taskSeq, arrivalTime, burstTime, number of tasks n;
 \ENDIF
  \FOR {$i = 1$ to n}
  \STATE select task i to be processed;
  \STATE dequeue the root element from $P_{queue}$
  \STATE scheduleSeq.append(i);
  \STATE $K^*$ are the available processors to process task i;
  \IF {($K^* \neq \emptyset $)}
  \STATE assign current task to $K$;
  \IF {(currentTime$<$arrivalTime[i])}
  \STATE currentTime = arrivalTime[i];
  \ENDIF
  \STATE startTime.append(currentTime);
  \STATE waitingTime = waitingTime + (currentTime-arrivalTime[i]);
  \STATE completionTime = currentTime + burstTime[i];
  \STATE currentTrunAroundTime = completionTime - arrivalTime[i];
  \STATE totalTurnAroundTime = totalTurnAroundTime + currentTrunAroundTime;
  \STATE turnAroundTime.append(currentTrunAroundTime);
  \STATE release $K$;
  \ELSE 
  \STATE return to if
  \ENDIF
  \ENDFOR
  \STATE calculate averageWaitingTime, averageTurnAroundTime; 
 \end{algorithmic} 
\end{algorithm}

\subsubsection{Multi-Task Hybrid Scheduling}
The incidents at the end of the priority queue need to wait longer when there is a large scale disaster because of plenty of rescue requests. Assume there are some tasks which need to wait longer for rescue due to lower priority. Suppose some of those tasks are located in an area where the disaster situation is worsening by time. The severity can increase fast at those places. A priority balancing scheduling policy might be helpful in such a scenario. It may need more information and human input to decide how and when to increase the priority of a task before it enters into critical condition. To solve this dilemma, we introduce a priority balancing module which re-calculate the priority score after the completion of each rescue mission. 

\begin{algorithm}[ht]
 \caption{Multi-tasks Hybrid Scheduling}
 \begin{algorithmic}[1]
 \renewcommand{\algorithmicrequire}{\textbf{Input:}}
 \renewcommand{\algorithmicensure}{\textbf{Output:}}
 \REQUIRE processorNo, taskSeq$[1...n]$, tasksPriority$[1...n]$, arrivalTime$[at_1...at_n]$, burstTime$[bt_1....bt_n]$, taskslocation[1...n], disRadius;
 \ENSURE  scheduleSeq$[task_i...task_n]$, startTime$[st_1...st_n]$, \\ turnAroundTime, averageWaitingTime, averageTurnAroundTime;
 \\ \textbf{Initialization}: All the processors K are released and ready to begin a task.\\ Initialize, scheduleSeq, startTime, and turnAroundTime as list []; currentTime = 0, waitingTime = 0, totalTurnAroundTime = 0;\\
Sort the variables in descending order using taskPriority. Re-sort the values in ascending order using burstTime and arrivalTime for same taskPriority tasks. Assign the tasks in priority queue $P_{queue}$;
\IF {(new task request)}
 \STATE update $P_{queue}$, and resort taskSeq, arrivalTime, burstTime, number of tasks n;
 \ENDIF
  \FOR {$i = 1$ to n}
  \STATE select task i to be processed;
  \STATE dequeue the root element from $P_{queue}$
  \STATE scheduleSeq.append(i);
  \STATE $K^*$ are the available processors to process task i;
  \IF {($K^* \neq \emptyset $ and available $K$ is capable of addressing task $i$)}
  \STATE assign current task to $K$;
  \FOR {$m = i+1$ to n}
  \STATE calculate the distance d of taskSeq[m] from current task using tasksLocation[m];
  \IF {(d$<$disRadius and $K$ has the extra resources to complete taskSeq[m] after current task)}
  \STATE add taskSeq[m] with the current task queue and create a sub-scheduling for those tasks;
  \STATE dequeue the taskSeq[m] and update $P_{queue}$;
  \ENDIF
  \ENDFOR
  \STATE estimate startTime, waitingTime, totalTurnAroundTime following the similar process of algorithm 2. 
  \STATE release $K$;
  \ELSE 
  \STATE return to if
  \ENDIF
  \ENDFOR
  \STATE calculate averageWaitingTime, averageTurnAroundTime; 
 \end{algorithmic} 
\end{algorithm}
Instead of a single rescue task in a mission, a rescue team can execute multiple tasks depending on available resources. For example, in a flood situation, several individuals can be rescued in the same boat and transferred to a shelter together. We illustrate this idea along with priority balancing in Algorithm 2. A processor can be assigned for multiple tasks in a single rescue mission if it has available resources. We use a 2 miles radius area for this purpose. A processor looks for other available tasks which are within 2 miles radius of the assigned event. It will incorporate multiple tasks as long as the processor has adequate resources and executes those tasks sequentially using priorities. Comparative performance evaluation of the algorithms is present in Section \ref{algoEval}. In Section \ref{realWorldAnalysis} we describe and demonstrate the  Multi-Task Hybrid Scheduling algorithm using a real-world disaster scenario.

\section{Experimental results}\label{er}
\subsection{Tweets Classifier Evaluation}\label{CE}
The primary goal of the tweet classification is to identify the people who need help and determine a priority score for each tweet based on the classified labels. To accomplish this goal, 4900 tweets were manually labeled into six different binary classes from 68,574 preprocessed tweets on Hurricane Harvey and Irma. We evaluate the proposed classification model on this labeled dataset and compared it with the well-established Logistic Regression (LR), Support Vector Machine (SVM) and fundamental CNN model. Moreover, in order to fully understand the effectiveness of our approach, we evaluate our model on several past disaster datasets obtained from CrisisNLP \cite{imran2016twitter} and CrisisLex \cite{olteanu2015expect}. We use the same datasets and data settings of Nguyen et al.\cite{nguyen2016rapid} and compare the output of our proposed model with the stated results of LR, SVM, and CNN in the same paper. Further, to evaluate the robustness of our proposed technique, we merged 15 different disasters data from CrisisLex \cite{olteanu2015expect} and perform a binary classification which identifies the tweets relevant to a particular disaster.

\subsubsection{\textbf{Model Parameters}} A set of optimal parameters is crucial to achieve desired performance results. We perform rigorous parameter tuning and select an optimal set that is used in all the experiments. We use the same parameter for better evaluation and model reproducibility. Table \ref{hyperparams} represents the used parameters for the model. The popular evaluation metrics such as precision, recall, F1-score, accuracy , and AUC score is used to validate and compare the experimental result of the models.




\subsubsection{\textbf{Evaluation on Hurricane Harvey and Hurricane Irma data}}
We use 4900 manually labeled tweets for this evaluation where 3920 tweets (80\%) used for training and the rest of the 20\% tweets used for testing. In the evaluation tables, we denote our model as $CNN_{AAf}$, which stands for CNN with Attention and Auxiliary features. We compare our model ($CNN_{AAf}$) with LR, SVM, and CNN without attention and auxiliary features. Our model outperformed all other models by more than 5\% inaccuracy metrics. In terms of precision, the proposed model performed surprisingly well and outperformed the closed result of SVM by around 25\%.

\begin{table}
\renewcommand{\arraystretch}{1}
\caption{Hyperparameter values}
\vspace{-1em}
\label{hyperparams}
\centering
\begin{tabular}{|p{2.25cm}|p{5.75cm}|}
\hline
\bfseries Hyperparameter & \bfseries Value/Description \\
\hline
Text embedding  & Dimension: 300 \\
\hline
 BLSTM Layer & 2 layers; 300 hidden units in each (Forward and Backward) \\
\hline
Conv1D Layer & 3 layers; 300 convolution filters \\
\hline
Dense Layer & 3 layers; First 2 layers have 150 and 75 units respectively and the last one is output (Dense) \\
\hline
Drop-out rate & Word Embedding: 0.3; Dense layer: 0.2 each;  \\
\hline
Activation function & Conv1D, BLSTM, Dense: ReLU; Output Dense layer: Sigmoid;  \\
\hline
Adam optimizer & Learning rate = 0.0001; $beta_1$=0.9; \\
\hline
Epochs and batch & Epochs = 10 to 25; batch size = 128; \\
\hline
\end{tabular}
\vspace{1.5em}
\end{table}

Table \ref{harveyIrma} represents the full evaluation results for the different classifiers. Table \ref{harveyIrmaClasses} represents the evaluation metrics for individual classes (Hurricane Harvey and Irma) using $CNN_{AAf}$ model. The distributions of the six classes in the data are Help - 29.1\%, Flood - 26.3\%, Water Needed - 4.9\%, DCEW - 4.1\%, Injured - 0.3\%, Sick - 0.3\%. However, we discarded labels Injured and Sick due to lack of enough data instances for training and testing so that it cannot influence the metrics of the model. As there are few true positive instances, those two labels achieve a higher rate of Accuracy although the model is not identifying true positive instances. We can also observe a better precision and accuracy for labels Water Needed and DCEW. This is happening as there are also a few true positive instances. However, still, the model has performed well for the recall and F1-score as the words found in the tweets for those labels have fewer variations. 

\begin{table}[ht]
\renewcommand{\arraystretch}{1}
\caption{Classifier evaluation (Hurricane Harvery and Irma)}
\vspace{-1em}
\label{harveyIrma}
\centering
\begin{tabular}{|p{2cm}|c|c|c|c|}
\hline
\bfseries Model  & \bfseries Precision & \bfseries Recall & \bfseries F1-score & \bfseries Accuracy\\
\hline
LR & 55.8 & 93.0 & 69.7 & 84.5\\
\hline
SVM & 65.1 & 85.4 & 73.9 & 88.5\\
\hline
CNN & 61.6 & 90.8 & 73.4 & 87.5\\
\hline
$CNN_{AAf}$ & 81.7 & 93.4 & 87.2 & 93.7\\
\hline
\end{tabular}
\end{table}

\begin{table}[ht]
\renewcommand{\arraystretch}{1}
\caption{Evaluation metrics for individual classes (Hurricane Harvery and Irma) using $CNN_{AAf}$ model}
\vspace{-1em}
\label{harveyIrmaClasses}
\centering
\begin{tabular}{|p{2cm}|c|c|c|c|}
\hline
\bfseries Class  & \bfseries Precision & \bfseries Recall & \bfseries F1-score & \bfseries Accuracy\\
\hline
Help & 87.9 & 97.7 & 91.2 & 94.9\\
\hline
Flood & 78.2 & 94.1 & 85.3 & 91.3\\
\hline
Water Needed & 87.5 & 71.4 & 78.7 & 98.0\\
\hline
DCEW & 93.7 & 73.2 & 82.3 & 98.5\\
\hline
Weighted Avg & 81.7 & 93.4 & 87.2 & 93.7\\
\hline
\end{tabular}
\end{table}

\subsubsection{\textbf{Evaluation on CrisisNLP and CrisisLex datasets}}
We use the same datasets and class distributions consisting of Nepal Earthquake, California Earthquake, Typhoon Hagupit, and Cyclone PAM which is described in \cite{imran2016twitter}. In that paper, the authors evaluate the models on event data, out-of-event data and a combination of both datasets. In table \ref{CrisisNLP}, we represent the results on the combination of both datasets. Clearly, our proposed $CNN_{AAf}$ model outperformed all other models in term of AUC score which the authors also used in the referenced paper. Auxiliary features have a high impact to better understand the semantic meaning of the tweets which is reflected on the AUC score. 

\begin{table}[ht]
\renewcommand{\arraystretch}{1}
\caption{Classifier evaluation AUC scores (CrisisNLP)}
\vspace{-1em}
\label{CrisisNLP}
\centering
\begin{tabular}{|p{3.5cm}|c|c|c|c|}
\hline
\bfseries Disaster Name  & \bfseries LR & \bfseries SVM & \bfseries \textbf{$CNN_I$} & \bfseries \textbf{$CNN_{AAf}$}\\
\hline
Nepal Earthquake & 82.6 & 83.6 & 84.8 & 87.5\\
\hline
California Earthquake & 75.5 & 74.7 & 78.3 & 83.6\\
\hline
Typhoon Hagupit & 75.9 & 77.64 & 85.8 & 88.3\\
\hline
Cyclone PAM & 90.6 & 90.74 & 92.6 & 92.6\\
\hline
\end{tabular}
\end{table}

\begin{table}[H]
\renewcommand{\arraystretch}{1}
\caption{Used Datasets from CrisisLex}
\label{crisisData}
\vspace{-1em}
\centering
\begin{tabular}{|p{8cm}|}
\hline
2012 Colorado wildfires, 2012 Costa Rica earthquake, 2012 Guatemala earthquake, 2012 Italy earthquakes, 2012 Philipinnes floods, 2012 Typhoon Pablo, 2012 Venezuela refinery, 2013 Alberta floods, 2013 Australia bushfire, 2013 Bohol earthquake, 2013 Colorado floods, 2013 Manila floods, 2013 Queensland floods, 2013 Sardinia floods, and 2013 Typhoon Yolanda. \\
\hline
\end{tabular}
\end{table}

We consider 15 different natural disaster datasets from CrisisLex \cite{olteanu2015expect} is presented in table \ref{crisisData}. After removing null values and preprocessing the merged datasets contains 13738 data instances. We use around 75\% data for training (9268) and validation (1030) and 25\% data for testing (3440). The comparative evaluation result using sklearn metrics is presented in table \ref{CrisisLex}. It is observable that the domain-specific auxiliary features along with attention layer is highly beneficial for understanding and identifying crisis tweets. Our proposed approach can be used on a diverse set of datasets with good outcome and this might play a crucial role to develop quick response application on the disaster domain. 
\begin{table}[ht]
\renewcommand{\arraystretch}{1}
\caption{Classifier evaluation  (CrisisLex)}
\vspace{-1em}
\label{CrisisLex}
\centering
\begin{tabular}{|p{2cm}|c|c|c|c|}
\hline
\bfseries Model  & \bfseries Precision & \bfseries Recall & \bfseries F1-score & \bfseries Accuracy\\
\hline
LR & 85.8 & 71.1 & 77.8 & 85.8\\
\hline
SVM & 90.9 & 74.7 & 82.1 & 73.2\\
\hline
CNN & 93.4 & 76.3 & 84.2 & 76.4\\
\hline
$CNN_{AAf}$ & 93.6 & 93.7 & 93.4 & 93.6\\
\hline
\end{tabular}
\end{table}


\subsection{Computational experiment on scheduling algorithms}\label{algoEval}
\begin{figure}[ht]
\centerline{\includegraphics[width=3.5in]{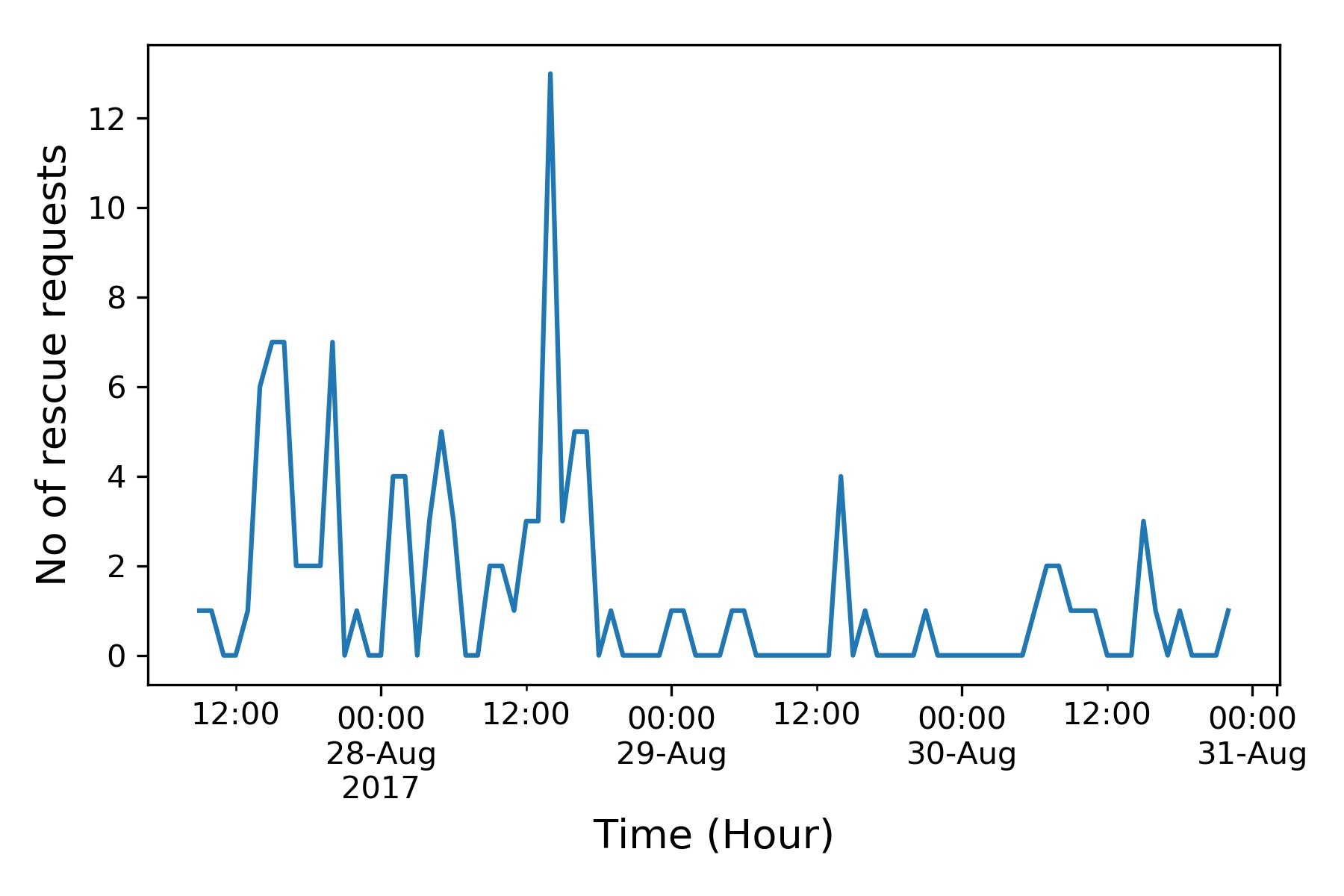}}
\vspace{-2em}
\caption{Distribution of rescue seeking tweets per hour}
\label{rescueReq}
\end{figure}

A computational experiment has been performed on the proposed algorithm in Section \ref{RSalgos}. For the purpose of evaluation and comparison, a data-set consisting of hurricane Harvey tweets between $27^{th}$ August 2017 and $31^{st}$ August 2017 have been used. To identify the rescue seeking tweets, the proposed tweet classification model is used. We processed the identified tweets to extract and determine the required information for the scheduling algorithm such as location, possible service time (burst time), and priority using the described process in Section \ref{dataColPro}. The priority of each tweet was determined on a scale of 10 using four classes (Flood, Water Needed, DCEW, and Sick or Injured), labeled by the classifier following Equation \ref{priority_func}. The weights for those classes were assigned as 1.5, 1.5, 2 and 2.5, respectively. For the environmental feature vector, we use a random distribution between 0.5 to 2.5. However, automatic weight determination still remains an open problem for the research. Next, the probable service time was estimated for each of the rescue tasks. We use the normal distribution of average service time as 54 minutes which is described in \cite{yang2017harvey}. Finally, after all the processing, 174 rescue seeking tweets were found from around 72 hours data frame. This sample size is relatively small and distributed over a longer period which is depicted in Figure \ref{rescueReq}. Hence, we performed upsampling using resample and linear interpolation methods from python pandas library \cite{mckinney2014pandas} and created a dataset containing 550 rescue tasks to evaluate the rescue algorithms.

\begin{figure}
\centerline{\includegraphics[width=3.5in]{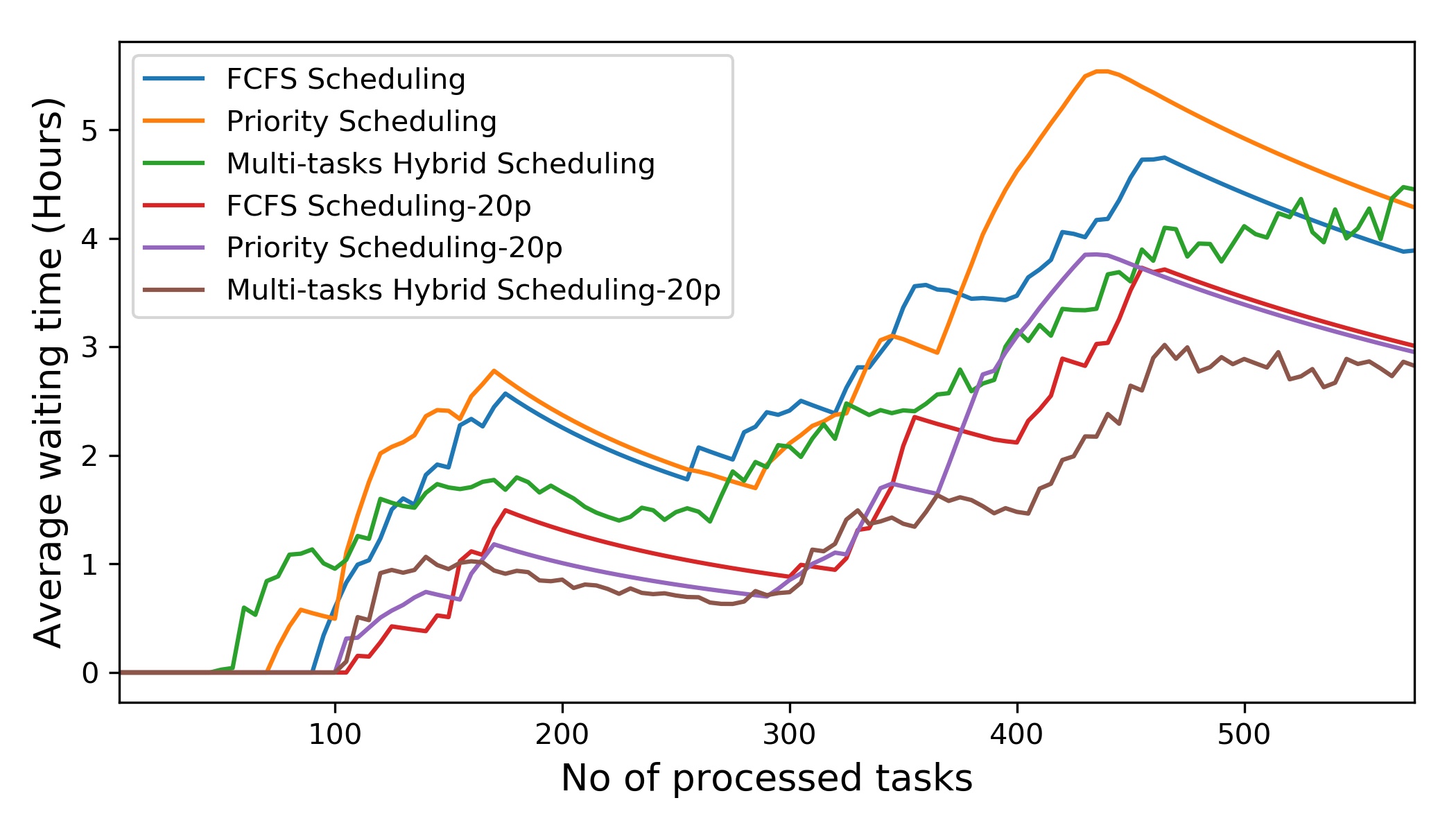}}
\vspace{-1em}
\caption{Average waiting time using 10 and 20 processors}
\vspace{0.5em}
\label{avgwt1020}
\end{figure}

The algorithms were implemented using the multiprocessing system. We use the number of rescue units (processors) as 10 and 20 to evaluate the performance of the scheduling algorithms. In Multi-task hybrid scheduling algorithm, the traveling time from one rescue location to another also considered while combining multiple tasks. Eventually, this estimation reduces the processing time for those tasks. Table \ref{wttable} describes the summary of the three algorithms. In the table, 10p and 20p represent the number of processors used to execute those algorithms. The average waiting times are lowest in case of Multi-task hybrid scheduling algorithm. The average waiting time (hours) with the number of processed tasks is represented in Figure \ref{avgwt1020}. The experimental results can be summarized as follows.

\begin{table}[ht]
\renewcommand{\arraystretch}{1}
\caption{Average waiting time summary}
\vspace{-1em}
\label{wttable}
\begin{center}
\begin{tabular}{| c | c | c | c | c |}
\hline
\textbf{Algorithms} & \multicolumn{2}{ c |}{\textbf{Max avg WT}}  & \multicolumn{2}{ c |}{\textbf{Mean avg WT}}\\ 
\cline{2-5}
 & \textbf{10p} & \textbf{20p} & \textbf{10p} & \textbf{20p} \\
\hline
FCFS & 4.74 & 3.73 & 2.53 & 1.61  \\ \hline
Priority & 5.54 & 3.85 & 2.81 & 1.63 \\ \hline
Multi-tasks Hybrid & 4.47 & 3.02 & 2.24 & 1.31 \\ \hline
\end{tabular}
\end{center}
\end{table}

\begin{itemize}
\item FCFS scheduling algorithm performs better comparing to Priority scheduling algorithm. However, in a disaster scenario, FCFS is not a fair policy to distribute the resources and rescue mission. Priority scheduling has a longer average waiting time because the lower priority tasks are waiting longer in the queue.

\item Multi-tasks hybrid scheduling beats all other algorithm with respect to average waiting time. This algorithm is more practical for effective rescue scheduling and resource allocation as it consider resource constraints. It allows completing small tasks together of a nearest distance. Furthermore, it can be utilized to transfer the required resources (such as water, medicine) to the different locations while optimizing the average waiting time. However, the maximum average waiting time for this algorithm can be high for a task with less priority and larger processing time. It can happen when the location of a mission is far away with a low priority score. 
\end{itemize} 

\subsection{Experimental analysis on real-world disaster scenario}\label{realWorldAnalysis}
\begin{figure}[ht]
\centerline{\includegraphics[width=3.2in]{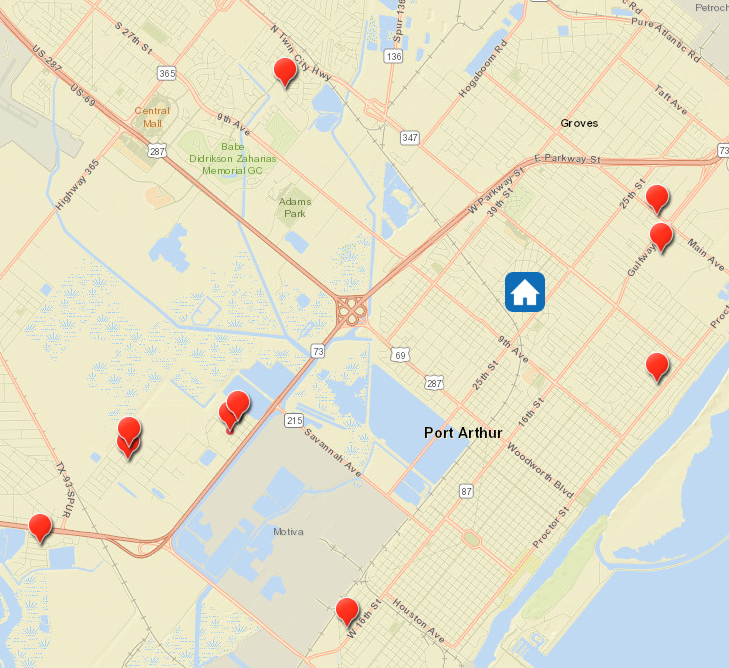}}
\caption{The geographical positions of the victims (Red icons) and hypothetical rescue operation base (Home icon).}
\label{portArthurMap}
\vspace{2em}
\end{figure}

\begin{figure}
\centerline{\includegraphics[width=3.5in]{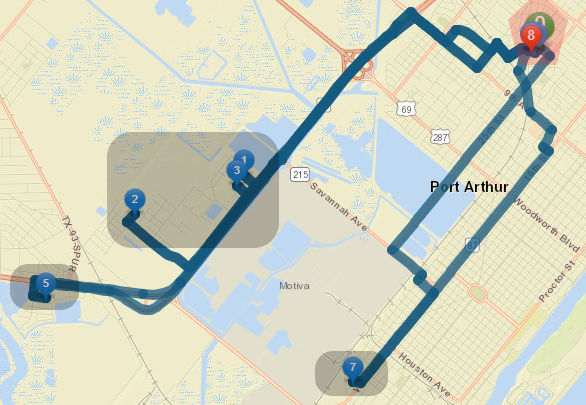}}
\vspace{-0.5em}
\caption{Route of Rescue Unit 1 by rescue order}
\label{rescuePaths}
\vspace{1em}
\end{figure}

A sample data-set is processed from the tweets during Hurricane Harvey to demonstrate Multi-task Hybrid Scheduling algorithm. An area of 20 square miles radius at Port Arthur, Texas has been selected for performing rescue operations. Figure \ref{portArthurMap} represents the geographical locations of the victims and the hyphothetical rescue operation base in the Port Arthur, Texas during hurricane Harvey. The ArcGIS javascript API \cite{arcgis} is used to create Figure \ref{portArthurMap} and \ref{rescuePaths}. To demonstrate the algorithm, we assume that there is a rescue operation base at Tyrrell Elementary School, Port Arthur, TX.  We simulate the rescue operations using 2 and 4 rescue units. The experimental process can be summarized by the following steps:

\begin{itemize} 
\item First, We have selected the rescue seeking tweets and extracted the location using the Stanford Named Entity Recognizer (NER) \cite{finkel2005incorporating} and Google map API. 

\item Second, we extracted 10 tweets which were arrived first and located around 20 miles radius of the rescue operation base after 12pm of 30th August 2017. 

\item Third, the priority score, probable burst time and distance metrics have been calculated for each of the 10 rescue tasks. 

\item Finally, the Multi-task Hybrid Scheduling algorithm created the rescue schedule. We have simulated the experiment using 2 and 4 rescue units and two different distributions of the possible burst time. First, we assumed the required burst time to be 54 minutes for each task based on the paper on hurricane Harvey rescue by Yang et al. \cite{yang2017harvey}. Further, we use a random completion time for the first 5 tasks and predict the burst time of future rescue missions using equation \ref{burstTimePred}.
\end{itemize} 

Tables \ref{tweetsLabelSample} and \ref{environmentVector} represent example data sample of tweet labels and environmental features for priority calculation using Equation \ref{priority_func}. We have used demo weights for the labels and environmental features as
(Flood - 1.5, Water Needed - 1.5, DCEW - 2, Sick or Injured - 2.5, Storm - 1, Road Damaged - 1, forecasted storm - 0.5, forecasted flood - 0.5) respectively. We used experimental weights because determining the weights for those labels and features requires domain expert and extensive study. An appropriate authority or domain expert will be able to input precise weight values for the labels and environmental features considering the situation
during an actual disaster. In the tables, $id$ represent the respective tweet which is later refers to the same numbered $taskSeq$ in Table \ref{portArthurData}. The calculated priorities also presented in Table \ref{portArthurData} as Priority Score. 

\begin{table}[ht]
\setlength{\tabcolsep}{0.5em}
\renewcommand{\arraystretch}{1}
\caption{Classified tweet labels for priority determination}
\vspace{-1em}
\label{tweetsLabelSample}
\centering
\begin{tabular}{|c|c|c|c|c|}
\hline
\bfseries id & 
\bfseries \begin{tabular}{@{}c@{}}Flood \\ \end{tabular}  & 
\bfseries \begin{tabular}{@{}c@{}}Water \\ Needed\end{tabular}  & 
\bfseries \begin{tabular}{@{}c@{}}DCEW \\ \end{tabular} & 
\bfseries \begin{tabular}{@{}c@{}}Sick or \\ Injured\end{tabular} \\
\hline
1 & 1 & 1 & 0 & 1\\
\hline
2 & 1 & 0 & 0 & 0\\
\hline
3 & 0 & 1 & 1 & 0\\
\hline
4 & 1 & 0 & 0 & 1\\
\hline
5 & 0 & 0 & 0 & 0\\
\hline
\end{tabular}
\end{table}
\vspace{-1em}

\begin{table}[ht]
\setlength{\tabcolsep}{0.5em}
\renewcommand{\arraystretch}{1}
\caption{Environmental features example}
\vspace{-1em}
\label{environmentVector}
\begin{center}
\begin{tabular}{| c | c | c | c | c |}
\hline
\textbf{id} & \multicolumn{2}{ c |}{\textbf{Current}}  & \multicolumn{2}{ c |}{\textbf{Forecast}}\\ 
\cline{2-5}
 & \textbf{Storm} & \textbf{Road Damaged} & \textbf{Storm} & \textbf{Flood} \\
\hline
1 & 0 & 1 & 1 & 0  \\ \hline
2 & 0 & 0 & 1 & 0 \\ \hline
3 & 0 & 1 & 0 & 1 \\ \hline
4 & 0 & 1 & 0 & 0 \\ \hline
5 & 1 & 0 & 0 & 0 \\ \hline
\end{tabular}
\end{center}
\end{table}

Table \ref{portArthurData} represents some columns of the processed sample data set of Port Arthur for the rescue scheduling. In the table, the burstTime is represented in minutes and distanceFromBase is measured in miles. To use Multi-task Hybrid Scheduling algorithm on the data, we need to assume some parameters. We consider the starting time of rescue mission as 14:00, the speed of the used vehicles or boats to rescue is 20MPH, and after the completion of each rescue mission a rescue unit requires 30 minutes as a preparation time before next task. 

\begin{table}[ht]
\setlength{\tabcolsep}{0.3em}
\renewcommand{\arraystretch}{1}
\caption{Real-world data sample for simulation}
\vspace{-1em}
\label{portArthurData}
\centering
\begin{tabular}{|c|c|c|c|c|}
\hline
\bfseries taskSeq & 
\bfseries \begin{tabular}{@{}c@{}}Arrival \\ Time\end{tabular}  & 
\bfseries \begin{tabular}{@{}c@{}}Burst \\ Time\end{tabular}  & 
\bfseries \begin{tabular}{@{}c@{}}Priority \\ Score\end{tabular} & 
\bfseries \begin{tabular}{@{}c@{}}Distance \\ from Base\end{tabular} \\
\hline
1 & 12:13 & 54 & 7 & 5.1\\
\hline
2 & 12:45 & 54 & 2 & 5.0\\
\hline
3 & 12:58 & 54 & 5 & 6.9\\
\hline
4 & 14:07 & 54 & 5 & 7.0\\
\hline
5 & 14:46 & 54 & 1 & 3.9\\
\hline
6 & 15:23 & 75 & 2 & 4.5\\
\hline
7 & 16:10 & 70 & 8 & 1.9\\
\hline
8 & 16:52 & 30 & 7 & 7.7\\
\hline
9 & 17:30 & 35 & 5 & 1.8\\
\hline
10 & 18:05 & 45 & 6 & 2.0\\
\hline
\end{tabular}
\vspace{1em}
\end{table}

The Multi-task Hybrid Scheduling algorithm can be demonstrated on the data in the table \ref{portArthurData} as follows. We use 2 rescue units to illustrate the algorithm. 

\begin{enumerate}
    \item The Start time of the rescue operation is 14:00. So, there will be 3 tasks in the queue at the time of the first iteration. The algorithm will first sort the tasks based on the priority score. Hence, the sorted sequence will be $1>=3>=2$.
    
    \item The location of the highest priority task (taskSeq 1) will be the point of interest. The algorithm will consider a perimeter of 2 square miles of that point and check if any other rescue task is there which can be combined. We can observe that taskSeq 1,2, and 3 are within 2 miles radius. If a rescue unit contains enough resource for running those 3 operations sequentially, it will combine those tasks and rescue the people in a single go without coming back and forth to the base. 
    
    \item The algorithm will further create a sub-schedule of 3 tasks assigned to rescue unit 1. Task 1 has the highest priority score and hence, the rescue unit will first go to location 1. From Figure \ref{rescuePaths}, we can observe that tasks 1 and 3 are in a close distance. However, task 2 has a higher priority. As the algorithm emphasizes the priority score most, it will schedule task 2 before task 3. The rescue unit will assist the people in location 2 and then come back to location 3. Finally, it will come back to the base after the completion of all 3 tasks.   
    
    \item  If there are multiple tasks with the high priority ($priority>=7$), separate rescue unit will be assigned despite of there occurrence in a close proximity. In our experimental setup if two tasks with high priority are within 2 miles radius, the algorithm assigns two separate units for those two tasks. However, if there is only one rescue unit available, the algorithm will follow the above approach. Multiple tasks with the same priority will be sorted based on burst time and arrival time, respectively. Multiple tasks with same priority and burst time will be sorted using arrival time.  
    
    \item Based on the conditions, taskSeq 1, 2 and 3 will be assigned to rescue unit 1. Rescue Unit 2 will take care of taskSeq 4 which arrives at 14:07. The algorithm will wait until the completion of a task, after which a rescue unit became available. 
    
    \item The taskSeq 4 will complete first and rescue unit 2 will become available around 16:13. The algorithm will iterate again and sort the remaining tasks. At this point, the queue contains 3 tasks (taskSeq 5,6 and 7).
    
    \item Employing the conditions, the sorted order for the tasks will be $7>=6>=5$. The taskSeq 7 has a high priority and there are no other victims nearby. Hence, rescue unit 2 will be assigned to complete task 7. 
    
    \item Rescue unit 1 will be available again at 17:55. The algorithm will continue iterating until all of the 10 tasks are completed. 
\end{enumerate}

\vspace{-0.5em}
\begin{table}[ht]
\setlength{\tabcolsep}{0.25em}
\renewcommand{\arraystretch}{1}
\caption{Rescue scheduling output table of Multi-tasks Hybrid Scheduling algorithm using 2 rescue units}
\vspace{-1em}
\label{partialOutput}
\centering
\begin{tabular}{|c|c|c|c|c|c|c|}
\hline
\bfseries taskSeq & 
\bfseries \begin{tabular}{@{}c@{}}Start \\ Time\end{tabular}  & 
\bfseries \begin{tabular}{@{}c@{}}Route \\ Distance\end{tabular} & 
\bfseries \begin{tabular}{@{}c@{}}Route \\ Duration\end{tabular} & 
\bfseries \begin{tabular}{@{}c@{}}Waiting \\ Time\end{tabular} & 
\bfseries \begin{tabular}{@{}c@{}}TAround \\ Time\end{tabular} &
\bfseries Unit\\
\hline
1 & 14:00 & 5.1 & 15 & 122 & 176 & 1\\
\hline
3 & 15:09 & 2.0 & 06 & 137 & 191 & 1\\
\hline
2 & 16:09 & 2.2 & 07 & 211 & 265 & 1\\
\hline
4 & 14:07 & 7.0 & 21 & 21 & 75 & 2\\
\hline
7 & 16:13 & 1.9 & 06 & 09 & 79 & 2\\
\hline
8 & 17:55 & 7.7 & 23 & 86 & 116 & 1\\
\hline
10 & 18:05 & 2.0 & 06 & 06 & 51 & 2\\
\hline
9 & 19:32 & 1.8 & 06 & 128 & 163 & 2\\
\hline
6 & 19:41 & 4.5 & 14 & 272 & 347 & 1\\
\hline
5 & 20:49 & 3.9 & 12 & 375 & 429 & 2\\
\hline
\end{tabular}
\end{table}

Table \ref{partialOutput} represents some output values and rescue schedule for the data illustrated in Table \ref{portArthurData}. The column $Start Time$ represents the scheduled time for the respective task. $Route Distance$ denotes the actual one-way path that a rescue unit needs to travel for a particular rescue mission. When multiple tasks are group together for a single mission the $Route Distance$ became the path between previous task and current task. For example, in Table \ref{portArthurData}, the distance of rescue location of taskSeq 3 from base is 6.9 miles. However, as tasks 1,2 and 3 grouped together the distance between the previous task 1 and task 3 became 2 miles. $Route Duration$ is the rounded time in minutes to travel the specific $Route Distance$. In our experiment, we assume that a rescue unit needs 3 minutes to travel a mile. $Waiting Time$ is the subtraction of $Start Time$ and $Arrival Time$  with the addition of required travel time ($Route Duration$) for a rescue location. The turnaround time is represented by $TAround Time$ in the table which is the summation of $Waiting Time$ and $Burst Time$. The rightmost column in the table represents the assigned rescue unit for a task. After returning from a rescue mission to the base, a rescue unit requires a preparation time to become available for the next mission. In the experiment above, the rescue unit 1 reached at the base at 17:25 after completing the first rescue mission of task 1,2 and 3. However, it became available at 17:55 after taking necessary preparations. 

The routes of the rescue missions assigned to rescue unit 1 presented in table \ref{partialOutput} are illustrated in Figure \ref{rescuePaths}. The red pentagon shadow area denotes the rescue operation base. The black shadowed rectangular shapes represent the rescue mission. Location points 1, 2, 3, 5 and 7 denote the taskSeq 1,3, 2, 8 and 6, respectively. Pointers 1,2, and 3 are inscribed in the same box as those tasks were combined together and performed in a single mission. The rescue unit 1 will start from the base (0) and travel to point 1, 2 and 3 to rescue victims and complete the tasks 1,3 and 2 in the first rescue mission. It will return back to base which is denoted by blue pointer (4) below the red pointer indicating 8. The unit will again travel to location 5, return to the base (6) and complete the taskSeq 8. Finally, the location of the third rescue mission pointed by 7 and the missions will be completed by rescue unit 1  after reaching to the base (point 8). 

We have also conducted the same experiment with 4 rescue units. The average waiting time and turnaround time reduced dramatically in this scenario. In the first experiment with 2 rescue units, the average waiting time and turnaround time is around 137 minutes and 189 minutes respectively. With 4 rescue units, waiting time and turnaround time came down to 49 minutes and 102 minutes. With the low number of rescue units, the tasks with low priority need to wait longer which increase the average waiting time. From table \ref{portArthurData} and table \ref{partialOutput}, we can observe that taskSeq 5 arrived at 14:46 with a priority score of 1. Due to the very low priority, task 5 scheduled last at time 20:49 with a waiting time of 375 minutes. However, tasks with higher priority such as 1, 7 and 8 had to wait a fairly lower amount of time.

\section{Conclusion and Future Work}
In this paper, we utilized social media (Twitter) for disaster management applications such as categorizing, identifying, and prioritizing users who need help and developed an algorithm for rescue scheduling. We introduced a novel approach for an effective rescue scheduling algorithm. First, we developed a tweet classifier using deep learning with attention layer and auxiliary features. The classifier labels every tweet into six different classes. Those labels allow us to identify the necessary information to assist the person/people in the tweet and estimate a priority score for that task. Second, we developed a multi-task hybrid scheduling algorithm and conducted the experiments using real disasters data for evaluating the efficiency of the algorithm. In the future, we would like to work on precise location determination and optimal estimation of the required time for a rescue mission. In addition, we are developing a fully-featured web application for deploying on the real-time disaster to evaluate the effectiveness of our work in disaster management. 

%
\begin{acks}
This research has been partially supported by a grant from NSF CNS-1461914.
\end{acks}

%
\bibliographystyle{ACM-Reference-Format}
\bibliography{sample-base}

%
\appendix

\end{document}